\documentclass[fleqn,usenatbib]{mnras}

\usepackage{newtxtext,newtxmath}
\usepackage[T1]{fontenc}

\DeclareRobustCommand{\VAN}[3]{#2}
\let\VANthebibliography\thebibliography
\def\thebibliography{\DeclareRobustCommand{\VAN}[3]{##3}\VANthebibliography}

\usepackage{bm}
\usepackage{graphicx}	% Including figure files
\usepackage{amsmath}	% Advanced maths commands
\usepackage{romannum}

\title{Constraints on Dark Matter-Neutrino Interaction from 21-cm Cosmology and Forecasts on SKA1-Low}

\author[A. Dey et al.]{Antara Dey,$^{1}$\thanks{E-mail: antaraaddey@gmail.com}
Arnab Paul$^{1,\, 2}$\thanks{E-mail: arnabpaul9292@gmail.com}
and Supratik Pal$^{1,\,3}$\thanks{E-mail: supratik@isical.ac.in}
\\
$^{1}$Physics and Applied Mathematics Unit, Indian Statistical Institute, Kolkata-700108, India\\
$^{2}$School of Physical Sciences, Indian Association for the Cultivation of Science, Kolkata-700032, India\\
$^{3}$Technology Innovation Hub on Data Science, Big Data Analytics and Data Curation,
	Indian Statistical Institute, Kolkata-700108, India\\}

\begin{document}
\label{firstpage}
\pagerange{\pageref{firstpage}--\pageref{lastpage}}
\maketitle

\begin{abstract}
In this article, we have done a thorough investigation of the possible effects of interaction between dark matter (DM) and neutrinos on reionization history. We have constrained the interaction strength using 21 cm Cosmology and found out possible deviations from standard, non-interacting $\Lambda$CDM scenario.  
Comparing the results with the existing constraints from present cosmological observations reveals that  21 cm observations are more competent to constrain the interaction strength by a few orders of magnitude. We have also searched for prospects of detecting any such interaction in the upcoming 21 cm mission SKA1-Low by doing a forecast analysis and error estimation.

\end{abstract}

\begin{keywords}
cosmology: dark matter, dark ages, reionization, first stars
\end{keywords}

%%%%%%%%%%%%%%%%%%%%%%%%%%%%%%%%%%%%%%%%%%%%%%%%%%

%%%%%%%%%%%%%%%%% BODY OF PAPER %%%%%%%%%%%%%%%%%%

\section{Introduction}
\label{Intro}

The standard model of Cosmology, the $\Lambda$CDM model, is so far the most widely accepted theory of the Universe, thanks to its compatibility with the highly precise data of Cosmic Microwave Background (CMB) anisotropy, Large Scale Structure (LSS) along with many other observations, barring couple of moderate to high tensions. In its vanilla version, the Universe is composed of a cosmological constant $\Lambda$ that takes into account  $\sim 70 \%$ of cosmic budget and $\sim 26 \%$ is attributed to cold dark matter (CDM). The CDM, assumed to be non-relativistic (hence dubbed as ''cold") and mostly non-interacting (other than the  gravitational interactions), conforms with a number of astrophysical and cosmological observations at all scales starting from galaxies \citep[]{1970ApJ...159..379R},  cluster of galaxies \citep[]{Randall_2008}, to large scale structures \citep[]{Tegmark_2004} and CMB \citep[]{aghanim2020planck}. 
In order to find out the exact constituent of DM, one needs to propose particle candidates. However, in the framework of particle physics, it is unnatural to incorporate DM  without (non-gravitational) interactions. Thus one of the major objectives of dark matter research is to search for its possible interactions, either with self or with any other species, either in terrestrial experiments or in astrophysical/cosmological observations.

Several well-motivated Beyond Standard Model (BSM) scenarios give rise to possible dark matter species with interactions with the Standard Model (SM) particles. Of them, a widely accepted candidate is weakly interacting massive particle (WIMP), in which the WIMP DM particles are in thermal equilibrium with the SM particles in the early Universe, leading to %its imprints in the early Universe by annihilation of the DM particle to SM particles which 
energy injection into the SM thermal bath due to DM annihilations, and the effect is observable in the CMB anisotropy spectrum \citep[]{Chluba_2010}, \citep[]{Finkbeiner_2012}, \citep[]{Galli_2013}, \citep[]{Slatyer_2009}. 
Scattering between DM and baryons leave imprint in the cosmological observables, hence have been constrained both in the context of CMB \citep[]{Boddy_2018} and 21 cm Cosmology \citep[]{Barkana_2018}. Other interesting possibilities include DM-DR (dark radiation) interactions \citep[]{Cyr_Racine_2016}, \citep[]{Escudero:2018thh} and DM-neutrino interactions, which have been studied in the light of CMB Power Spectrum \citep[]{Mangano_2006},  \citep[]{Wilkinson_2014}, \citep[]{Escudero_2015}, \citep[]{Escudero_2018}, \citep[]{Di_Valentino_2018}, \citep[]{Olivares_Del_Campo_2018}, \citep[]{Stadler_2019}, \citep[]{Ghosh:2019tab}, \citep[]{Mosbech:2020ahp}, \citep[]{Paul:2021ewd}. DM annihilation into highly energetic neutrinos have also well studied in the literature \citep[]{Blennow_2008} constraining DM annihilation cross-section at current epoch \citep[]{Arg_elles_2021}, \citep[]{frankiewicz2015searching}.

Among the above-mentioned possibilities, we consider the interaction between DM and neutrinos and explore its possible  signatures in reionization history.
For the early stages of the Universe, when the perturbation theory is valid, the evolution of the fluctuations of DM and neutrino fluids can be studied using the first order Boltzmann equations, hence estimating its impact in CMB anisotropy spectrum. This have been exhaustively studied in the literature as mentioned before.

On cosmological scales, sizable DM-neutrino interactions may significantly impact the structure formation of the Universe, resulting from the modified evolution of DM over-density. This non-standard interaction provides with a  pressure-like term in the DM fluid, thus preventing DM from collapsing significantly at small length scales. Similar to the acoustic waves in the baryon-photon fluid, a dark acoustic oscillation along with the collisional and silk damping-like phenomenon is also observed in this case. Although the imprint of this oscillation may be observed in the linear matter power spectrum, non-linear evolution at late time washes out some of these features, leaving only a suppression of power at small scales. The suppression of the power at small scales may hence significantly reduce the halo formation and number of stars and galaxies in those halos, resulting in a smaller number of ionizing photons in the intergalactic medium (IGM). This lower number of ionizing photons reduce the amount of ionized hydrogen in the IGM and consequently, affects the reionization history. 

This is the primary target of the present article. We have made a thorough investigation of the possible effects of  DM-neutrino interaction  on reionization history. In the process, we have constrained the interaction strength using 21 cm Cosmology. We found that the constraints on DM-neutrino interaction from 21 cm missions is much more stringent compared to the constraints from existing cosmological observations. We also find a non-trivial effect of DM-neutrino interaction in HI maps. We have done forecast analysis in the
upcoming 21 cm mission Square Kilometer Array 1 - Low (SKA1-Low).

The paper is organized as follows: in section \ref{boltzman} we have discussed about the cosmological scalar perturbation equations in presence of DM- neutrino scattering, in section \ref{Method} we describe the semi-numerical reionization simulation and constraints of these parameter from different observations, in \ref{Results} constraints on the DM-neutrino parameter from reionization physics is discussed and detection techniques of this interaction are presented in \ref{error} for upcoming telescope SKA$1$-Low.

Throughout this work, we have adopted cosmological parameters $\Omega_\text{m}=0.3183, \Omega_\Lambda=0.6817, h=0.6704, \Omega_\text{b}h^2=0.022032, \sigma_8=0.8347, n_\text{s}=0.9619$ consistent with Planck+WP best-fit values \citep[]{aghanim2020planck}.

\section{Effect of DM-neutrino Interaction}\label{boltzman}
%\section[The EoR C II \texorpdfstring{158$\umu$m}{} and H I 21 cm line intensity mapping signals]{The E\lowercase{o}R C\,{\sevensize II} \texorpdfstring{158$\bmath{\umu}$\lowercase{m}}{} and H\,{\sevensize I} 21-\lowercase{cm} line intensity mapping signals} \label{Signal}

In order to consider DM-neutrino interaction one needs to go beyond the six parameter description of vanilla $\Lambda$CDM model and analyze the new set of perturbation equations. Let us describe the present situation briefly.
Here we will work on Newtonian gauge for our analysis. Presence of scattering between two cosmic species (DM and neutrino in the present case) results in a drag term in the perturbed Euler equation for the corresponding species.  Assuming a flat Universe, the modified perturbation equations for dark matter and neutrino fluid, in presence of DM-neutrino interaction is given by \citep[]{Stadler_2019},

\begin{eqnarray}
   \dot{\delta}_{\nu} &=& -\dfrac{4}{3} \theta_{\nu} + 4 \dot{\phi} \\
	 \dot{\theta_{\nu}} &=& k^{2}\Psi + k^{2} \left(\dfrac{1}{4}\delta_{\nu}-\sigma_{\nu}\right) - \dot{\mu}(\theta_{\nu}-\theta_{DM}) \\
\dot{\delta}_{DM} &=& - \theta_{DM} + 3 \dot{\phi} \\
  \dot{\theta}_{DM} &=& k^{2}\Psi - \mathcal{H}\theta_{DM} -S^{-1}\dot{\mu}(\theta_{DM}-\theta_{\nu}) 
\end{eqnarray}	
 	
\noindent where  an overdot represents a derivative with respect to conformal time. Here $\delta_{\nu}$, $\delta_{DM}$ are respectively the neutrino and DM density contrast and $\theta_{\nu}$, $\theta_{DM}$ corresponds to their velocity divergences; $k$ is the comoving wavenumber, $\Psi$ and $\Phi$ are the metric perturbations, $\sigma_{\nu}$ is the anisotropic stress potential and $\mathcal{H}$ is the conformal Hubble parameter. Further, the DM-neutrino interaction rate is expressed as $\dot{\mu}\equiv a \sigma_{DM-\nu}cn_{DM}$ where $\sigma_{DM-\nu}cn_{DM}$ is DM-neutrino elastic scattering cross-section and $n_{DM}$ is number density of DM. We have considered the neutrinos to be massless in our analysis for computational simplicity. For convenience, let us introduce a dimensionless quantity $u$ to parametrize the DM-neutrino interaction strength,

\begin{equation}
u\equiv \dfrac{\sigma_{DM-\nu}}{\sigma_{TH}} \left(\dfrac{m_{DM}}{100~\rm GeV}\right)^{-1}
\end{equation}
In principle, $\sigma_{DM-\nu}$ can be velocity dependent \citep[]{Wilkinson_2014}. However, in the present analysis we will  consider $\sigma_{DM-\nu}$ to be constant.%, without losing any essential information as such. 
This dimensionless quantity $u$ will act as a new parameter on top of the 6-parameter in the present scenario.

Over the years, the constraint on $u$ have improved considerably with the availability of new datasets. For example, using Planck 2013 and WP measurement the constraints on u was found to be $ u< 3.27 \times 10^{-2}$ \citep[]{Wilkinson_2014}. Recent studies using Planck 2018 data-set (high-$\ell$ TT+TE+EE, low-$\ell$ TT, low-$\ell$ EE) the constraint comes out to be a tighter $ u< 3.34 \times 10^{-4}$ \citep[]{Mosbech:2020ahp}, \citep[]{Paul:2021ewd}, the scattering cross-section constraint being $\sigma_{DM-\nu} < 7.44 \times 10^{-29}  cm^{2}$. As mentioned earlier, in this article, our primary intention is to investigate if 21 cm Cosmology can put further constraints on the parameter $u$ and the prospects of SKA1-Low in this regard, which will in turn help us have better idea on more stringent constraints on DM-neutrino interaction from future observations.

\begin{figure}
    \includegraphics[width=\columnwidth]{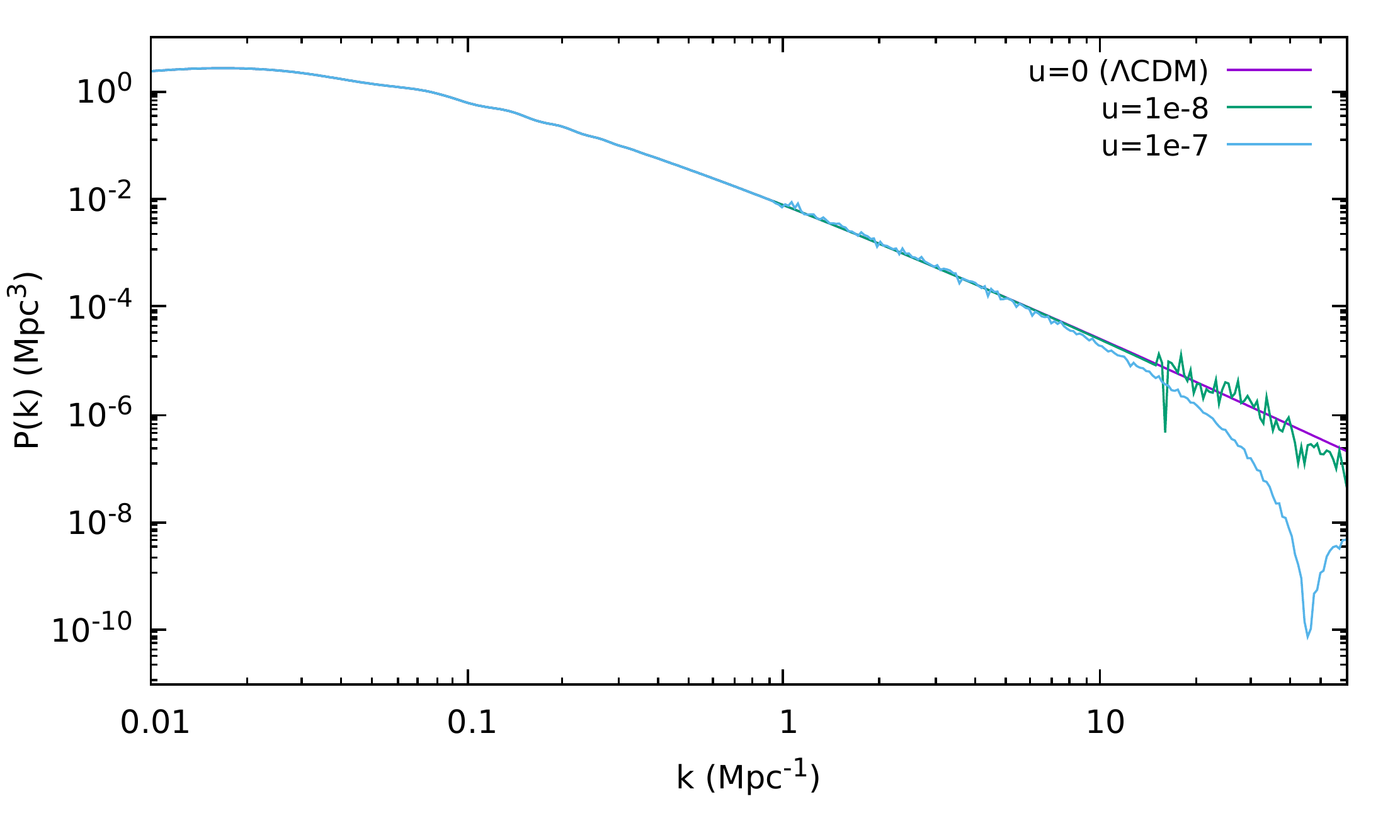}
	\caption{\textbf{Effects of $u$ in linear matter power spectrum:} This figure shows suppression and oscillatory feature in the presence of DM-neutrino interaction where $u\equiv [\sigma_{DM-\nu} / \sigma_{TH}] [m_{DM} / {100GeV}]^{-1}$  where $u=0.0$ corresponds to $\Lambda$CDM case and $u=10^{-8}, 10^{-7}$ signifies different strengths of interaction. We observe that, suppression in power at small scales increases with $u$.
	}
\label{fig1}
\end{figure}

Let us first find out the effects of DM-neutrino interaction on the matter power spectrum in Fig. \ref{fig1} using a modified version \citep[]{Stadler_2019} of the CLASS code \citep[]{Blas_2011}. We have illustrated this for specific values of the interaction parameter $u$ irrespective of observational constraints. For this, we have considered flat $\Lambda$CDM model with DM-neutrino interaction on top of it, where the other cosmological parameters are taken as the Planck 2018 best-fit values \citep[]{aghanim2020planck}. Fig. \ref{fig1} reveals that the  effects of DM-neutrino interaction (with $u< 10^{-4}$) on the matter power spectrum is being reflected by  a series of damped oscillations with a suppression of power in the non-linear regime ($k\ge 0.2 h Mpc^{-1}$). 

Having convinced ourselves on the non-trivial effects of DM-neutrino interaction, we are now in a position to explore more realistic situation, namely, the reionization simulation, corresponding HI maps and error estimation for SKA1-Low, that will help us drawing important conclusions on the prospects of 21 cm cosmology in this regard. We will do that stepwise in the subsequent sections.

%In the literature DM-neutrino \apaul{keep notation consistent, use DM-$\nu$ everywhere} scattering cross-section is constrained using Planck 2013 data \cite{2014} $\sigma_{DM-\nu} <= 10^{-33} (m_{DM}/GeV) cm^{2}$. 
%Recent studies with Planck 2018 data (high-l TT+TE+EE, low-l TT, low-l EE) puts robust constraints on scattering cross-section $\sigma_{DM-\nu} < 6.75 \times 10^{-29}  cm^{2}$ \cite{Paul:2021ewd}.
   
\section{Methodology}
\label{Method}

\subsection{Reionization Simulation}
 Cosmic reionization is one of the most crucial epochs of the evolution history of the Universe when the hydrogen content in the Universe gets ionized from its neutral state. The high energy radiation from the first luminous sources  starts  ionizing the neutral hydrogen in the intergalactic medium (IGM) in its vicinity. As reionization period progresses the ionized bubble gets bigger and starts to overlap until it fills the entire IGM. At the present epoch, IGM is almost completely in the ionized state, except a small fraction found in the dense core of galaxies in the form of a damped Lyman-$\alpha$ forest. Free electrons in the IGM interact with the cosmic microwave background radiation through Thomson scattering and optical depth $\tau_{\rm Th}$ corresponding to the scattering can be measured from the CMB observations, hence providing information about the reionization epoch. However, the onset and completion of the epoch of reionization depends on  the model of reionization and may vary from model to model. Latest constraints on the Thomson scattering optical depth is $\tau_{Th}$ = 0.054 ± 0.007 by taking into account different reionization model. This suggests that IGM is almost 10 percent ionized at redshift z$\sim$10 \citep[]{aghanim2020planck}.

In this article, we have used semi-numerical simulation proposed in \citep[]{Majumdar_2014}, \citep[]{Mondal_2015} to generate the 21 cm maps and the redshifted 21 cm brightness temperature power spectra at a redshift $z=8.0$. The three-step procedure for this is the following:
\begin{itemize}
    \item  The first step is to simulate the DM particle density distribution using the particle mesh algorithm \citep[]{Bharadwaj_2004}. DM particles have been simulated in a volume of $(150~\rm Mpc)^{3}$ with spatial resolution of 0.07 \rm Mpc and mass resolution of $1.09 \times 10^{8} M_{\rm Sun}$. The initial distribution for the DM particles are provided by the linear matter power spectrum mentioned in section \ref{boltzman}. The initial condition for the simulation is set at an earlier redshift at $z=124$ where initial density fluctuations are approximated as gaussian random field where $\Delta (k) = \sqrt{\dfrac{V P(k)}{2}}[a(k)+ib(k)]$; a(k) $\&$ b(k) are two real valued independent gaussian random variable of unit variance. Here $P(k)$ is the input linear matter power spectrum for interacting DM-neutrino scenario for different interaction strength. The initial particle position and velocities are estimated from Zel'dovich approximation $x \rightarrow x - \nabla \nabla^{-2} \delta(x)$, $v \rightarrow -aHf\nabla \nabla^{-2} \delta(x)$ where H is Hubble parameter and f ($=\frac{d lnD}{lna}$) logarithmic derivative of growth factor w.r.t. the scale factor. Then the DM particle distribution is being evaluated at lower redshifts by solving the equation of motion of each particle under the gravitational field produced by other particles in expanding Universe. And finally we get the particle position and velocity distribution at redshift of interest and the non-linear matter power spectrum at that redshift is being calculated from the DM density distribution. 
    
    \item The second step is to identify the halos using the halo finding technique via friends-of-friends (FOF) algorithm. We have taken the linking length as 0.2 times mean inter particle separation. We have set the criteria of halo to consist of a minimum of 10 dark matter particles which corresponds to minimum halo mass of $1.09 \times 10^{9}$ $M_{\rm Sun}$. Given the particle position and velocity distribution FOF algorithm counts the number of DM particles within a sphere of radius which is 0.2 times mean inter particle separation and if the number of particles inside this sphere reaches 10 it will then mark this region as a dark matter halo.
    
    \item The third, and final, step in this process is to generate the 21 cm HI map and differential brightness temperature power spectrum based on excursion set formalism \cite[]{Furlanetto_2004} for which we have used the semi-numerical code ReionYuga \citep[]{Majumdar_2014}, \citep[]{Mondal_2016}, based on \citep[]{Choudhury_2009}. 

\end{itemize}

 In the process, we have made certain assumptions while computing observables in the reionization era. We have considered only hydrogen ionization in the Universe, Helium ionization is not taken into account. Hydrogen gas follows underlying DM distribution and sources of the ionization are the ultraviolet radiations which are located inside the DM halo. 
 
 There are three parameters in our reionization model, namely, the minimum halo mass $M_{\rm min}$, ionizing photons per baryon $N_{\rm ion}$ and mean free path of the ionizing photons $R_{\rm mfp}$. Let us now discuss these three parameters in details.

\subsubsection{$M_{\rm min}$} This parameter represents the minimum mass of the halo that can host star formation. Once the the potential well of the DM halo is formed, baryons will simply fall into the well and as the gas tries to settle into the potential well, mergers will heat the gas to the virial temperature. In order to form galaxies, the gas has to dissipate its thermal energy and cool down. Depending upon the nature of the gas the cooling criteria will be different. For example, if the gas contains only atomic hydrogen it can never be cooled to lower than $10^{4}$K, whereas, if the gas contains molecules it can be effectively cooled to much lower temperature. However whether or not there are sufficient amount of molecules in the gas at high redshifts is yet to be known conclusively. The minimum of the halo mass will solely depend upon the cooling criteria of the baryons \citep[]{Barkana_2001},  \citep[]{10.1111/j.1365-2966.2004.07942.x}. The reionization simulation as done by \citep[]{Choudhury_2009} suggests that $M_{\rm min}$ has to be of the order of  $\approx 10^{6}-10^{7} M_{\rm Sun}$ to satisfy constraints from Thomson scattering optical depth and Gunn-Peterson troughs observations. But this analysis has not taken into account the Population $\Romannum{3}$ stars in the simulation which are an efficient source of ionizing radiation. The value of $M_{\rm min}$ is generally expected to be large if Population $\Romannum{3}$ stars are also included. Recent simulation by \citep[]{bouwens2017z} have constrained the value of $M_{\rm min}$ well within the allowed range of reionization $6< z <8$ using UV luminosity function data. We have chosen the value of $M_{\rm min}$ to be $1.09 \times 10^{9}$ $M_{\rm Sun}$ for our simulation \citep[]{Shaw_2020}.

\subsubsection{$N_{\rm ion}$} This is the number of photons entering into the IGM per baryon in collapsed objects. Our model assumes that ionizing UV photons from galaxies entering into the IGM is proportional to the halo mass $M_{h}$. The proportionality relation can be expressed as,
 \begin{equation}
 N_{\gamma} (M_{h}) = N_{\rm ion} \dfrac{M_{h}}{m_{p}} \dfrac{\Omega_{b}}{\Omega_{m}}
 \end{equation}
 
  Here $N_{\rm ion}$ is the dimensionless proportionality constant, $m_{p}$ is the proton mass and $\Omega_{b}$ $\&$ $\Omega_{m}$ are respectively the baryon and matter density parameter. This factor depends on several unknown parameters:  properties of ionizing sources, star forming efficiency $f_{*}$, escape fraction of ionizing photons $f_{\rm esc}$, hydrogen recombination rate \citep[]{Choudhury_2009}. In general $N_{\rm ion}$ can be dependent on the redshift, however in this work, we assume $N_{\rm ion}$ to be independent of redshift  for simplicity. We have chosen a fiducial value of $N_{\rm ion}$ to be 23.21. This is usually adopted for $\Lambda$CDM in order to achieve $50 \%$ ionization at redshift z = 8.0  satisfying the existing constrain on reionization 
  \citep[]{Choudhury_2009}. However, since we do not expect much deviation from $\Lambda$CDM anyway, this turns out to be a valid choice for our case as well.
  
  \subsubsection{$R_{\rm mfp}$} This factor depends upon the distribution of the Lyman-$\alpha$ systems in the IGM. From the observations of Lyman-$\alpha$  systems it has been observed that $R_{\rm mfp}$ can have values between $3-80 Mpc$ \citep[]{Songaila_2010} at z$\approx 6$. Accordingly, we have chosen the fiducial value of $R_{\rm mfp}$ to be $20 Mpc$ .
 
 The semi-numerical code we have used, is based on the excursion set formalism \cite[]{Furlanetto_2004}. A point on the grid $x$ is said to be fully ionized \textbf{i.e. $x_{i} =1 $} only if it satisfies the following condition
 $\langle n_{\gamma}(x)\rangle_{R}$ $  \ge $  $ \langle n_{H}(x)\rangle_{R}$
 where $\langle n_{\gamma}(x)\rangle_{R}$ is number density of ionizing photons smoothed over a sphere of comoving radius $R$ and $ \langle n_{H}(x)\rangle_{R}$ is the smoothed number density of hydrogen. And if the condition is not fully satisfied then the grid is said to be partially ionized with $x_{i} = \langle n_{\gamma}(x)\rangle_{R}/\langle n_{H}(x)\rangle_{R}$. The radius $R$ is varied from a minimum value i.e. grid size to maximum value of $R_{\rm mfp}$. The simulated H1 distribution mapped to redshift space using peculiar velocity of the particles in order to generate brightness temperature power spectrum. The final H1 maps are generated on a grid which is eight times coarser than the N-body simulations.
 
 As already stated, the fiducial values of the parameters under consideration [$M_{\rm min}, N_{\rm ion}, R_{\rm mfp}$] are [$1.08 \times 10^{9} M_{\rm Sun}$, 23.21, 20 Mpc] \citep[]{Shaw_2020}. Using these fiducial values, we have run the N-body code on 2144 grid size and generated the observables HI map and differential brightness temperature power spectrum.  

\subsection{Ionization Condition}

 In the analytical model of reionization, the ionization process can be expressed as \citep[]{choudhury2009analytical},

\begin{equation}
\dfrac{\mathrm{d}Q_{H\Romannum{2}}}{\mathrm{d}t} =\dfrac{\dot{n}_{\nu}}{n_{H}} -Q_{H\Romannum{2}}C \alpha(T) n_{H},
\end{equation}  
where $Q_{H\Romannum{2}}$ is volume filling factor and $n_{\gamma}$ is the source term which hosts the ionizing photons and term in right hand is responsible for the recombination, $C$ is clumping factor and $\alpha(T)$ is recombination rate coefficient at temperature $T$.
 
 In analytical and semi-numerical modelling of the reionization, there arise several parameters the values of which are not too precise. These uncertainties can be somewhat dealt with by the existing data from different sources, like quasar absorption spectra, Lyman-$\alpha$ data, CMB data etc. From one such modelling \citep[]{Mitra_2015} it has been shown that reionization process begins at around redshift $z=15$ and ends at $z=6$, but a huge range of $x_{H\Romannum{1}}$ is allowed in order to satisfy the current observational constraints. Since there are quite a good deal of uncertainties in the value of $x_{H\Romannum{1}}$, we have chosen a particular value $x_{H\Romannum{1}}=0.5$ for our model which is fairly justified considering all the existing observations taken together. Additionally, we have considered $50\%$ ionization condition at redshift $z=8.0$ which is also justified in order to complete the reionization process during the redshift range $z=\{13.0 - 6.0\}$, as in \citep[]{Mitra_2015}, \citep[]{Kulkarni:2018erh}.

\subsection{Constraints on $N_{\rm ion}$}
In order to find out possible constraints on $N_{\rm ion}$, we have assumed a linear relationship between the halo mass and number of ionizing photons emitted from that halo,
$N_{\gamma} (M_{h}) = N_{\rm ion} \dfrac{M_{h}}{m_{p}} \dfrac{\omega_{b}}{\omega_{m}}$ \citep[]{Choudhury_2009},
where $N_{\rm ion}$ is a proportionality factor as mentioned earlier. We can estimate this factor roughly as follows \citep[]{Mesinger_2016},

\begin{equation}
  N_{\rm ion} = 8  \dfrac{N_{\rm ion}^{b}}{4000}  \dfrac{M_{b}/M_{\rm halo}}{1/5} \dfrac{f_{*}}{10\%} \dfrac{f_{\rm esc}}{10\%}
\end{equation}
    
Here $N_{\rm ion}^{b}$ is the number of ionizing photons produced in the halo per neutral hydrogen (since we are ignoring helium in our case). $M_{b}/M_{\rm halo}$ is the baryonic mass fraction. %$f_{*}, f_{esc}$ are parameters, for which we are not certain about the values or dependencies.%the two factors which are unknown parameters, we do not know the values and exact dependencies of these parameters.
$f_{*}$ is the fraction of baryonic mass which has converted into stars and $f_{\rm esc}$ is the fraction of ionizing photons which has escaped from halo and enters into the IGM, this fraction is responsible for ionizing the neutral hydrogen in the IGM, however the values and exact dependencies of these parameters are uncertain. There is a huge uncertainty in the value of $N_{\rm ion}$ for population $\Romannum{3}$ (zero metallicity stars) and population $\Romannum{2}$ stars, due to large systematic uncertainties of $f_{\rm esc}$, $f_{*}$. For $Z=0.01$ metallicity stars, number of photons per baryon could be of the order of 4000, but for zero metallicity stars $N_{\rm ion}^{b}$ lie in the range $10^{4}-10^{5}$, where $N_{\rm ion}$ can be of $20$ to $200$ order. But these stars are short lived stars and the impact of these stars on reionization is considerably small. In general we can safely take $N_{\rm ion} \le 500 $ for population $\Romannum{2}$ stars \citep[]{10.1111/j.1365-2966.2004.07942.x} \citep[]{Barkana_2001} \citep[]{Choudhury_2009}.

\section{Results $\&$ Analysis}
\label{Results}
Let us now discuss the major results obtained from our analysis.

\begin{itemize}

 \item In order to find out the effect of DM-neutrino interaction on the reionization epoch, we have performed the semi-numerical simulation for different strength of interaction parameter `$u$' (as given in Table \ref{tableu}) and we have considered the same initial random seed for all the simulations. We have considered benchmark values for which neutral hydrogen fraction $x_{H\Romannum{1}}\sim0.5$ at redshift $z=8.0$.
 Introduction of the DM-neutrino interaction suppresses the matter power spectrum w.r.t. the $\Lambda$CDM model in the non-linear regime, thereby hampering dark matter halo formation. As a result, it will reduce star formation and lesser number of ionizing photons will be produced, hence delaying the reionization process. So, in order to achieve $50 \%$ ionization $z=8$ in the interacting dark scenario, $N_{\rm ion}$ has to be increased. However, for Population $\Romannum{2}$ stars,  $N_{\rm ion}$ beyond $500 $ is not physically allowed. So, we have restricted ourselves to $N_{\rm ion}<500$ condition. Keeping two reionization parameters $M_{\rm min}$ and $R_{\rm mfp}$ fixed at $1.08 \times 10^{9} M_{\rm Sun}$ and $20 Mpc$, we have varied the parameters $u$ and $N_{\rm ion}$ in order to achieve $x_{H\Romannum{1}}=0.5$. In order to make the analysis robust, we have taken  three
different values of $N_{\rm ion}$ and searched for possible range of the interaction parameter $u$ for which  $50 \%$ ionization condition can be achieved. In order to satisfy two conditions $N_{\rm ion} < 500$ and $x_{H\Romannum{1}}= 0.5$ simultaneously we have found that the the maximum value of the interaction parameter can be $u \leq 6.6 \times 10^{-7}$  that can be achieved from the reionization physics using these two criteria.\footnote{A somewhat similar analysis has been put forward in arXiv after this paper, that investigated for the bounds on '$u$' for massive neutrinos \citep[]{https://doi.org/10.48550/arxiv.2207.03107} that reassures our analysis.}

\item
It is noteworthy that the current constraints on the DM-neutrino interaction in light of different existing cosmological observations are not that stringent. For example, using Planck 2018 data (high-$\ell$ TT+TE+EE, low-$\ell$ TT, low-$\ell$ EE) the allowed range of the interaction parameter is of the order $u < 3.34 \times 10^{-4}$ \citep[]{Mosbech:2020ahp}, \citep[]{Paul:2021ewd}.
Our analysis reveals that putting 21 cm observation on top of existing observations results in a much more tighter constraint on the DM-neutrino interaction  ($u \leq 6.6 \times 10^{-7}$). So, the  upcoming 21 cm missions like SKA would have very good prospects to constrain DM-neutrino interaction further compared to existing observations and hence to improve our understanding about this cosmic entities. 
 
  \begin{table}
 	\begin{tabular}{l | c | c | }
 		DM-$\nu$ interaction Model (u) & $N_{\rm ion}$ & $x_{H\Romannum{1}}$ \\
 		\hline \hline

 	0.0 ($\Lambda$CDM) & 24 & 0.5  \\
 	%5.0 $ \times 10^{-10}$ & 24  & 0.51\\
 	8.8 $ \times 10^{-8}$ & 300 & 0.51 \\
 	6.6 $ \times10^{-7}$ & 500 & 0.51\\

	\end{tabular}
 		\caption{Table of $u$ (dark matter neutrino interaction strength), $N_{\rm ion}$ (number of photons entering IGM per baryon in collapsed objects) and $x_{H\Romannum{1}}$(neutral hydrogen fraction). Increase in the value of $u$ will suppress more power in the power spectrum. It will generate a lesser number of dark matter halo and less ionizing radiation will produce. In order to reach $50 \%$ ionization condition $N_{\rm ion}$ has to be increased. }
 		\label{tableu}
 \end{table}

\item We have done the N-body simulation for three different cases and generated a non-linear matter power spectrum (Fig. \ref{fig2}) at redshift $z=8.0$. From the figure, it is obvious that no oscillatory signature in the power spectrum is observed, as non-linearities washes out these features. Nevertheless, the maximum suppression of the power is observed for $u=6.6 \times 10^{-7}$. 
 \begin{figure}
	\includegraphics[width=\columnwidth]{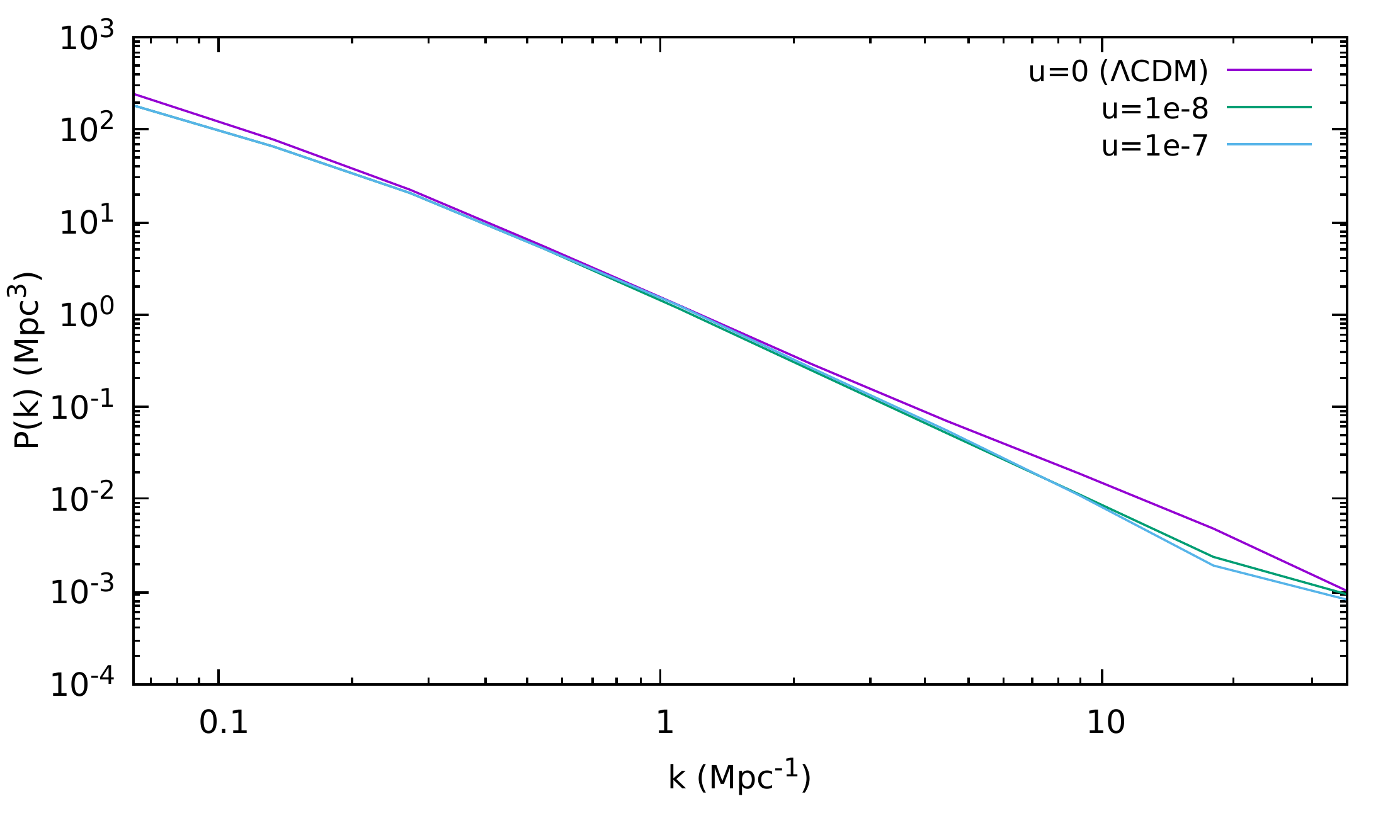}
	\caption{\textbf{Non-linear matter power spectrum:} The above figure shows the non-linear matter power spectrum at redshift $z=8.0$ calculated from N-body simulation for different strength of dark matter neutrino interaction $u$. We can see the suppression of power for increasing value of $u$, but the non-linearities washes out the oscillatory features of the power spectrum.}
\label{fig2}
\end{figure}

\item For different values of the interaction parameter $u$ and $N_{\rm ion}$,  different ionization features appear. As justified earlier, we have restricted our analysis to $x_{H\Romannum{1}}=0.5$ and $N_{\rm ion} < 500$ condition at $z=8.0$. However the assumption of $x_{H\Romannum{1}}=0.5$ at $z=8.0$ has considerable uncertainties, which may change the bounds on $u$ to some extent. With this we have performed  the reionization simulation and have generated HI maps for the above three representative value of $u$. These have been depicted in Fig. \ref{fig3}.  The maps reveal that increase in the value of $u$ within its allowed range decreases ionization and delays reionization physics. 
In order to achieve $50 \%$ ionization condition $N_{\rm ion}$ has to be increased. We observe slight increase in bubble size as $N_{\rm ion}$ increases, confirming the non-trivial effect of interaction on HI maps. Further, as u increases, we found slightly lower number of dark matter halos forming in the universe, as expected from the suppression of dark matter power spectrum. However, this change resulting from the figures may not be significant enough so as to quantify and find out its edge over standard $\Lambda$CDM scenario, mostly because of the feeble value of the interaction strength. 

\item We argue that our results are solely depend on the two constraints which are $x_{HI}=0.5$ at $z=8.0$ which is justified from Planck 2018 data and $N_{ion}<500$ for Pop-II stars. The two main observable quantities which constraints reionization epoch are the optical depth to reionization ($\tau_{Th}$ = 0.054 $\pm$ 0.007) and Gunn Peterson bound from quasar absorption spectra which tells abount the end of reionization. We have considered a conservative limit on $x_{HI}$ so that it can satisfy the two observational bounds.
However, if we consider the $50 \%$ ionization criteria at some smaller redshift \citep[]{Bouwens:2015vha}, \citep[]{Wang:2020zae}, \citep[]{Mason:2019ixe}, \citep[]{Villanueva-Domingo:2021vbi}, say at $z=7$, then it will affect  the constraints on $u$ and the conclusion may differ slightly from our case. While these aspects worth investigating, we do not expect a significant deviation from the conservative approach taken up in this article.

\begin{figure*}
\includegraphics[scale=0.46]{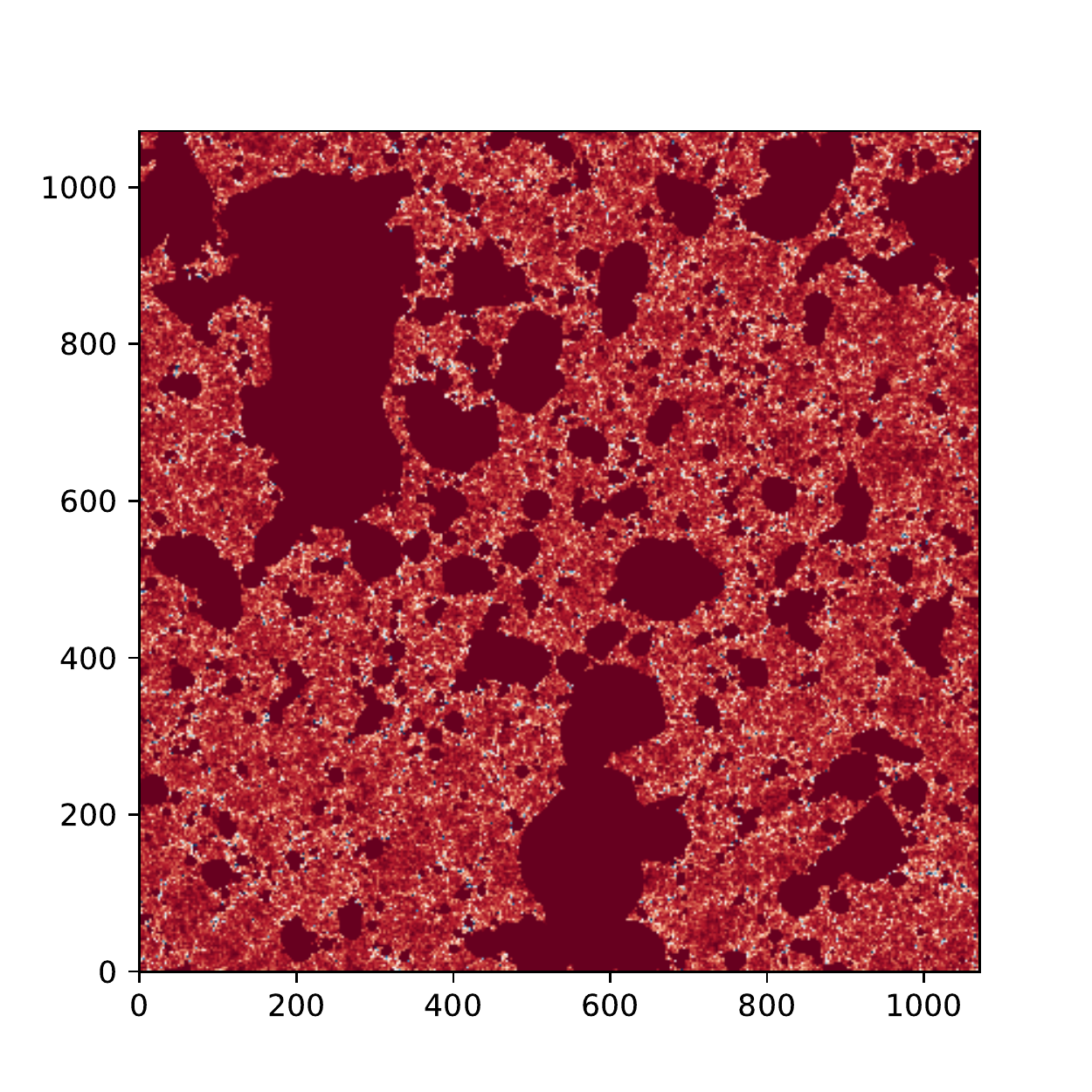}
\includegraphics[scale=0.46]{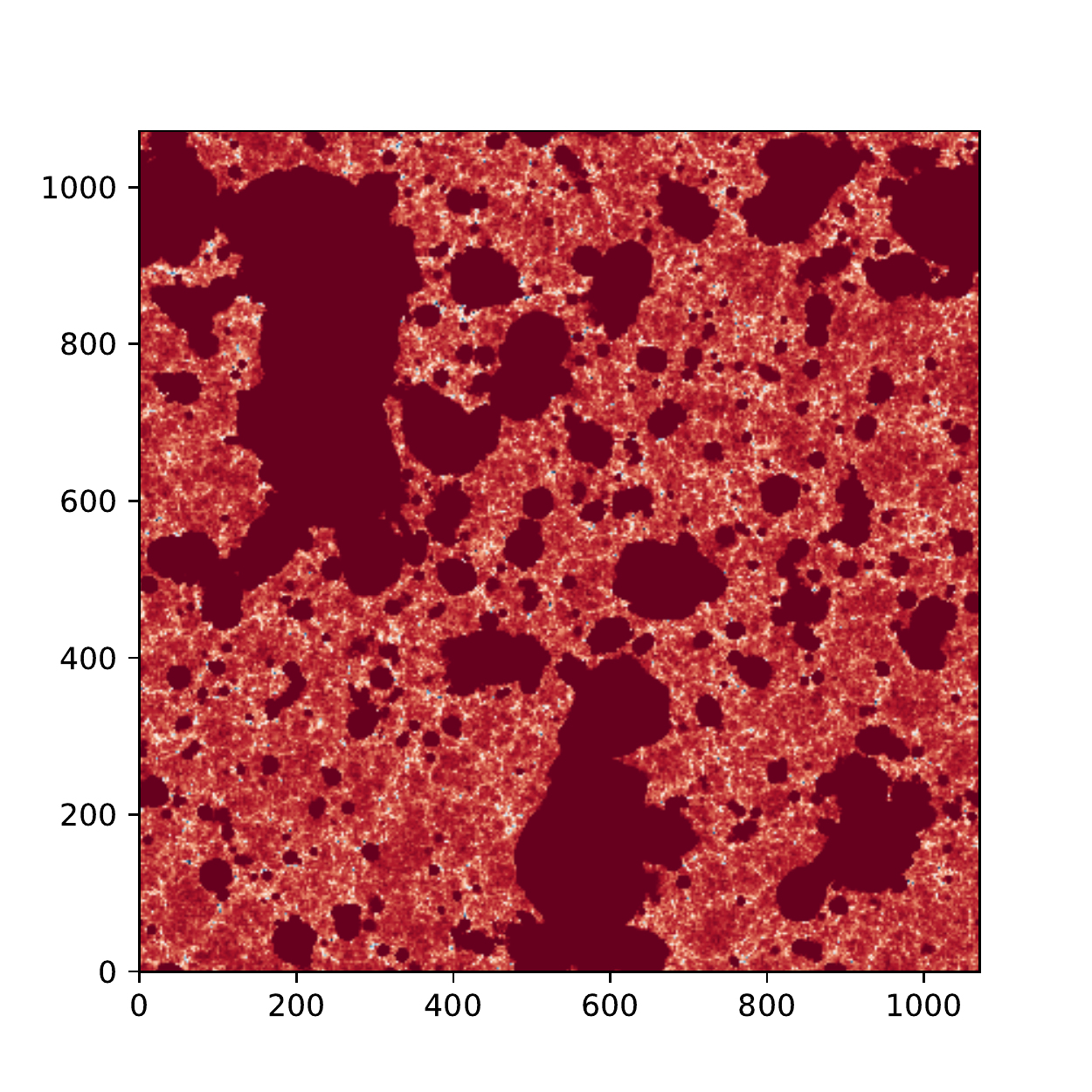}
\includegraphics[scale=0.46]{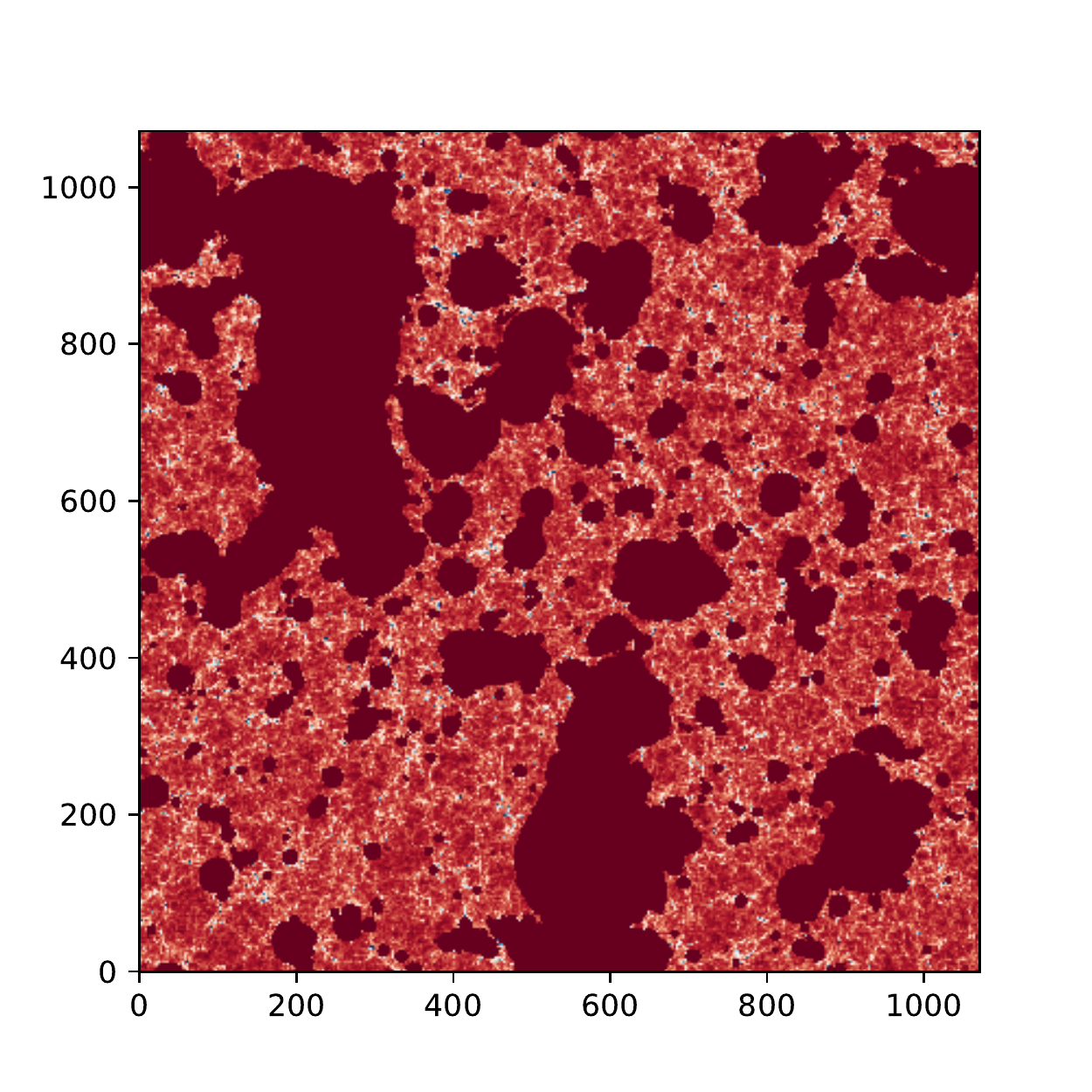}
\caption{\textbf{HI map:} Two dimensional section of the simulated HI map at $z=8.0$ generated from $2144^3$ grid size of the simulation box. The above figures are for $u=0$, $N_{\rm ion}=24$; $u=8.8 \times 10^{-8}$, $N_{\rm ion}=300$ and $u=6.6 \times 10^{-7}$, $N_{\rm ion}=500$, units along x and y axis are in $0.06~ \rm Mpc$. Darker shades corresponds to more ionized regions. For all the simulations we have assumed the identical initial random seed.} 
\label{fig3}
\end{figure*}

\end{itemize}

\section{Error Analysis $\&$ Prospects of Detection in SKA1-Low} 
\label{error}

Having convinced ourselves of the role of 21 cm observations in putting tighter constraints on DM-neutrino interaction, let us now investigate the prospects of 
the upcoming radio observations in this regard. To this end, we have done a  forecast analysis on the upcoming SKA1-Low telescope. The specifications for SKA1-Low are shown in Table \ref{ska1lowspecs}. Considering SKA1-Low antenna layout 
\citep[]{SKALowv2} 
we have simulated the distribution of antenna pair separations 'd'  corresponding to 8 hours of observations with an  integration time of 60 seconds towards a fixed sky direction located at DEC= $-30^{\circ}$. In addition to the signal i.e 21 cm brightness temperature fluctuation, observed 21 cm differential brightness temperature will have a contribution from random Gaussian system noise corresponding to each and every grid point in the simulation. The noise power spectra is given by \citep[]{10.1093/mnras/sty3242}, \citep[]{Shaw_2019}, \citep[]{Mondal_2020}. 

\begin{equation}
P_{N}(k_{g}) = \dfrac{8 ~\rm hours}{t_{\rm obs}} \dfrac{P_{0}}{\tau_{k_{g}}}
\end{equation}
where $P_{0}$ is the system noise power spectrum for a single visibility measurement with 60 seconds integration time. $\tau_{k_{g}}$ quantifies the number of independent visibility measurements.  Here $t_{\rm obs}$ is the observation performed over number of nights, considering a run in each night consists of 8 hours. So, effectively, the noise power spectrum will depend on the number of observation nights (in the denominator) scaled by 8 hours (in the numerator).

 \begin{table}
 	\begin{tabular}{l | c | }
 		Specifications & SKA$1$-LOW  \\
 		\hline \hline
 	Number of antennas & 131,072   \\ 

Collecting areas $m^{2}$ & 4,19,000 \\
 	Frequency coverage (MHz)  & 50-350  \\
 	Instantaneous Bandwidth (MHz) & 300 \\
 	Field of View ($deg^{2}$) & 2.3 - 113 \\
 	Maximum baseline (km) & 65 \\
 	Angular resolution (arcsec) & 3.3 - 23 \\
 	Sensitivity ($\mu Jy$ beam-1hr-1, fractional bandwidth 0.3) & 14 - 26
 
 	\end{tabular}
		\caption{Instrumental specifications for SKA$1$-Low telescope}
		\label{ska1lowspecs}
 \end{table}

We have compared the signal i.e. differential brightness temperature fluctuation with the noise power spectra for 1024, 10000, 50000 hours of observation and the results have been shown in Fig. \ref{fig6}. While computing the noise power spectrum we have considered the effect of cosmic variance. We can write the cosmic variance as \citep[]{Shaw_2019},
\begin{equation}
C_{ij} = \dfrac{\bar{P}(k_{i})^{2}}{N_{k_{i}}} \delta_{ij} + \dfrac{\Bar{T}(k_{i},k_{j})}{V}
\end{equation}
where $\bar{P}(k_{i})$ is bin averaged power spectrum and $N_{k_{i}}$ is number of independent k modes in each bin. In our analysis we have assumed 21 cm signal is completely Gaussian so we have neglected the trispectrum $(\Bar{T}(k_{i},k_{j}))$ contribution of the cosmic variance. If one incorporates the tri-spectrum contribution in the calculation then it will effect noise power spectrum slightly and the correlations which we have estimated for the astrophysical parameters will change slightly. However, if the signal is intrinsically non-gaussian at this redshift range, it should affect the calculations considerably. We hope to explore that direction in future.

We have assumed the foregrounds which plays an important role for detecting low frequency radio signal, is removed completely after modelling it so that the entire k space is allowed for detecting the signal. We have observed that in each of these cases the signal is higher w.r.t. the noise for all ranges of k, thereby giving a sizable SNR as required for detection. 

\begin{figure*}\textbf{
\includegraphics[scale=0.26]{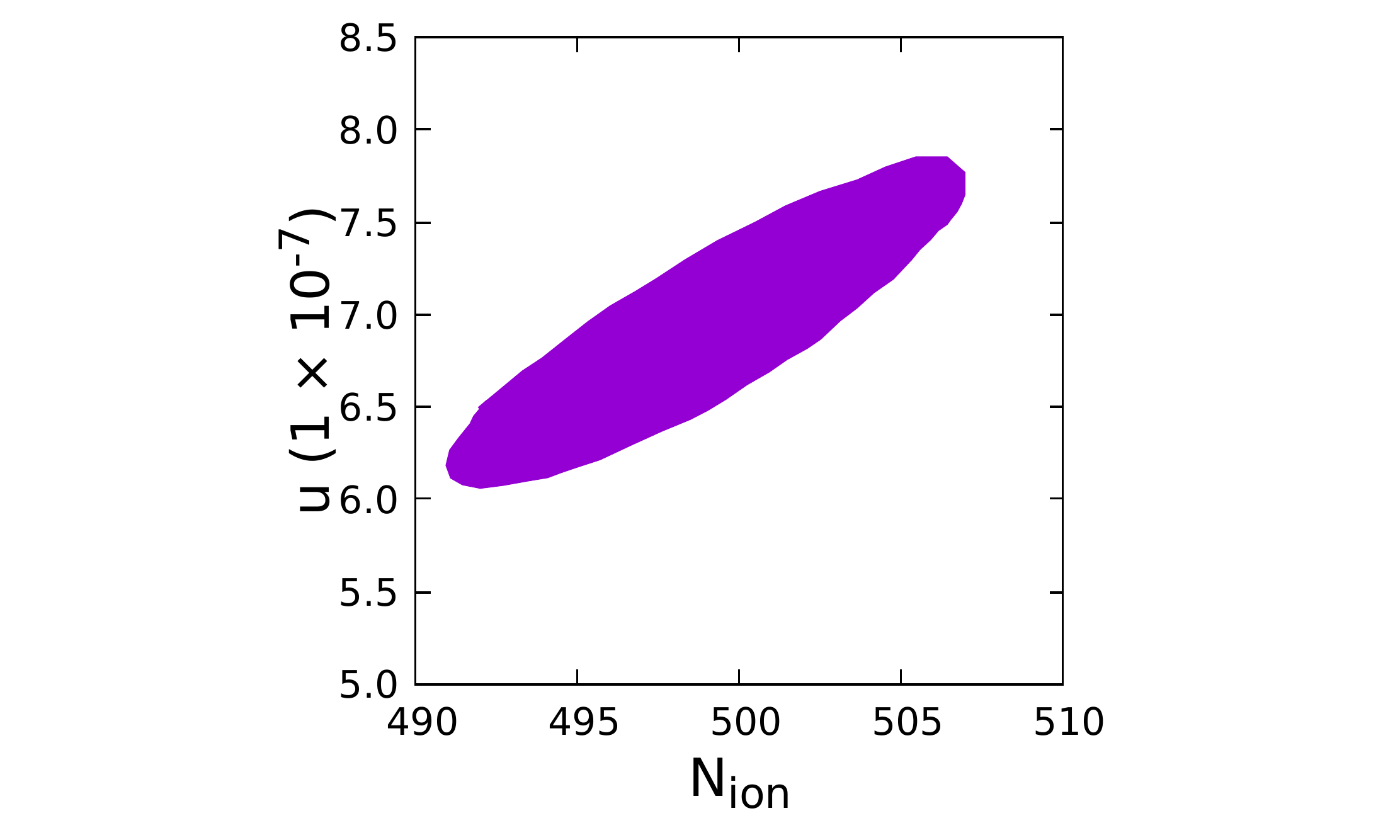}
\includegraphics[scale=0.26]{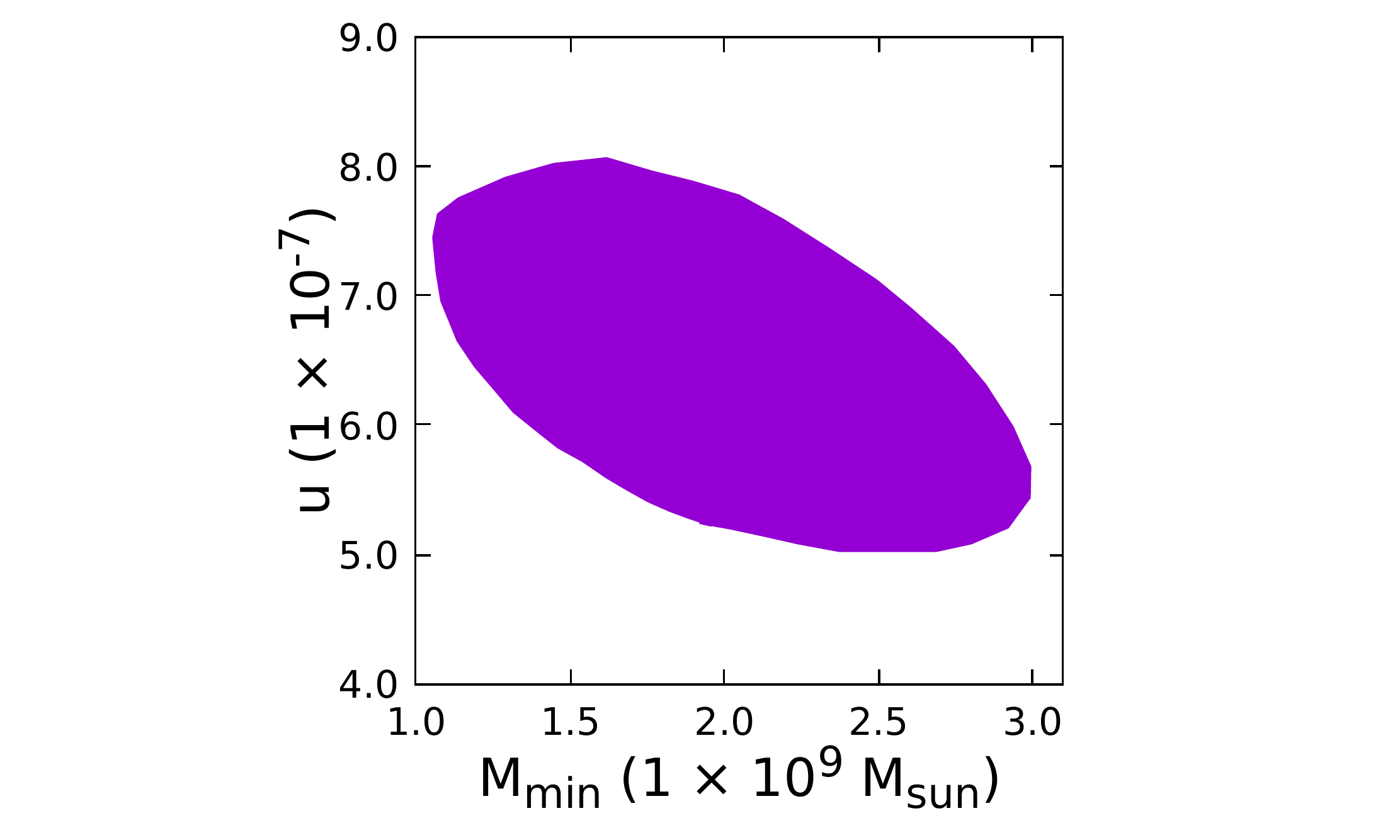}
\includegraphics[scale=0.26]{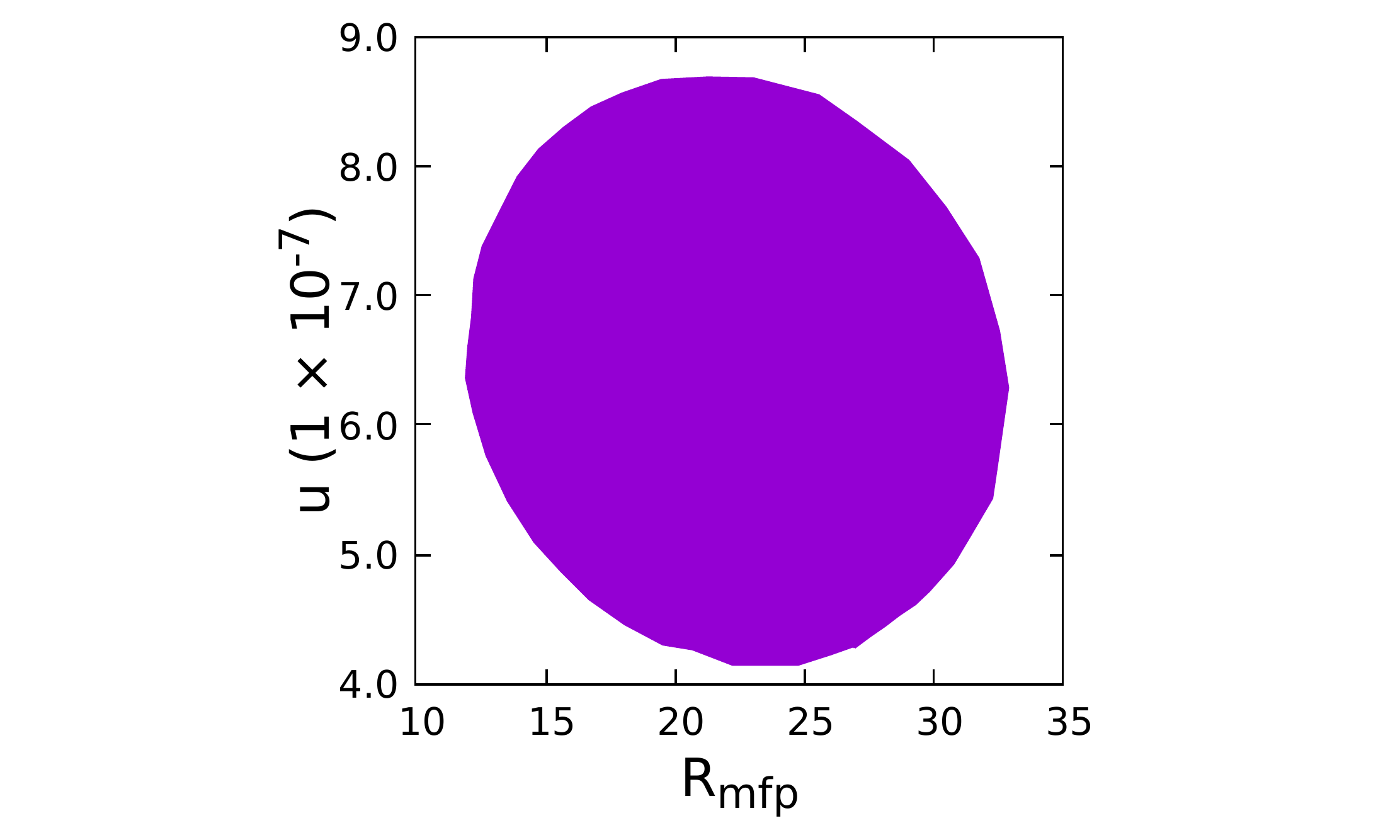}}
\caption{\textbf{1$\sigma$ error ellipse:} This plot represents the 1$\sigma$ error ellipse of interaction parameter u $\&$ $N_{\rm ion}$, u $\&$ $M_{\rm min}$ and u $\&$ $R_{\rm mfp}$ from Fisher matrix analysis.}
\label{fig5}
\end{figure*}

\begin{figure}
\includegraphics[width=\columnwidth]{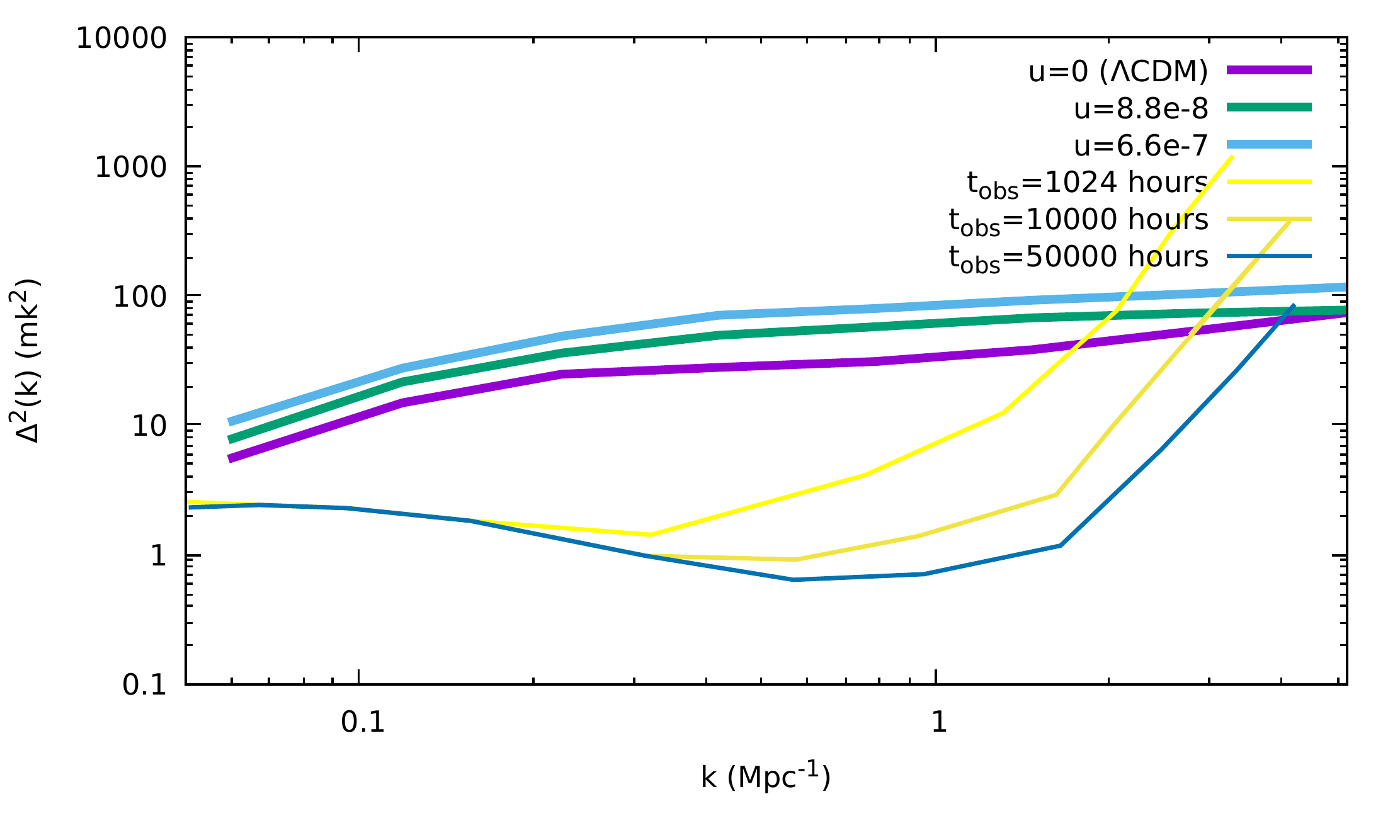} 
	\caption{\textbf{Comparison between Signal and Noise Power Spectra for SKA1-Low configuration:} The thick lines correspond to power spectrum of 21 cm brightness temperature fluctuations for u=0 ($\Lambda$CDM), $~8.8\times10^{-8}$ and $~6.6\times10^{-7}$ (darker shade denotes lower $u$) with $N_{\rm ion}=24,~300,~500$ correspondingly. The thin lines denote noise power spectrum generated from SKA1-Low proposed configuration for 1024, 10000, 50000 hours of observation (darker shade denotes higher observation time). The figure shows that the signal to noise ratio is large for minimum 1024 hours of observation. }
\label{fig6}
\end{figure}

We have done fisher forecast analysis for our astrophysical parameters and model parameter. The fisher matrix corresponding to the model parameter $q_{\alpha}$ is \citep[]{10.1093/mnras/stv2212}
\begin{equation}
F_{\alpha \beta} = \sum_{ij} (\dfrac{\partial  \bar{P}(k_{i})}{\partial q_{\alpha}} [C^{-1}]_{ij} \dfrac{\partial \bar{P}(k_{j})}{\partial q_{\beta}})
\end{equation}
where $C_{ij}$ is covariance matrix and $\Bar{P}(k_{i})$ is bin averaged power spectrum.
In Fig. \ref{fig5} the forecast on the parameter $u$, $N_{\rm ion}$, $M_{\rm min}$ and $R_{\rm mfp}$ has shown. From the analysis we have found a positive correlation on $u$ and $N_{\rm ion}$ and a negative correlation between $u$ and $M_{\rm min}$. Further from our analysis we found there is no significant dependencies on $R_{\rm mfp}$ and $u$.

In Fig. \ref{fig6} a comparison between Signal and Noise Power Spectra for SKA1-Low configuration has been demonstrated and explained in details. For $u>0$, the power at small length scales in both the linear and non-linear matter power spectra are suppressed, as a result less number of low mass halos are formed, resulting in lower number of ionizing sources. Hence the ionization at small length scales are suppressed, giving rise to more HI fraction, resulting an increase in the HI power spectrum.
An essential conclusion that comes out of the error analysis as in  Fig. \ref{fig6} is that the prospects of detection of any interaction between  DM and neutrino, if any, of strength as constrained by reionization epoch in the upcoming SKA1-Low telescope is quite high. So, this guarantees further investigation in this direction, some of which we hope to take up as we go along.

\section{Summary} \label{Summary}
In this work we have examined the effect of DM-neutrino interaction in the reionization epoch. We have run the semi-numerical N-body simulation to compute the reionization observables. We have put a constraint in the dark matter neutrino interaction parameter $u \equiv 6.6 \times 10^{-7}$ using $50 \%$ ionization criteria and $N_{\rm ion} \le 500$ condition at redshift $z=8.0$. Reionization puts stronger constraints to the interaction  parameter $u$ than any other observations studied in the literature. We also found a non-trivial effect of the said interaction on HI maps via a slight yet distinguishable change in the sizes of the bubbles. Further, we have done the forecast analysis for the upcoming SKA1-Low telescope and found that  the DM-neutrino interaction signal using this telescope can be observed  for a very small time of observation and we observe possible correlations of the reionization parameters with interaction parameter. Thus, future 21 cm mission like SKA1-Low can help us comment in a more convincing way whether or not these two cosmic species interact, and if so, what can be the possible interaction strength. This will not only shed some light on these entities on top of their properties as constrained by particle physical theory and experiments but also help us have better understanding of these cosmic species on top of the existing cosmological observations like CMB. 

The analysis done  in the present article mostly focuses on reionization era with typical redshift of our consideration $z=8$. However, if at all DM and neutrinos interact, that may reflect on post-reionization era as well. A thorough investigation of the effects of such interaction on post-reionization era needs to be done that will act as a complimentary to the constraints as obtained from 21 cm Physics during reionization. We look forward to explore these aspects in near future.

\section*{Acknowledgements}

Authors gratefully acknowledge the use of publicly available code \texttt{CLASS} 
\citep[]{Blas_2011}, \texttt{Particle Mesh N-body code} \citep[]{Bharadwaj_2004}, \texttt{Friends-of-Friends code} \citep[]{Mondal_2015}, \texttt{ReionYuga code} \citep[]{Choudhury_2009}, \citep[]{Majumdar_2014}, \citep[]{Mondal_2016} and thank the computational facilities of Indian Statistical Institute, Kolkata. AD thanks ISI Kolkata for financial support through Senior Research Fellowship and Abinash Kumar Shaw for useful discussions.   
SP thanks the Department of Science and Technology, Govt. of India for partial support through Grant No. NMICPS/006/MD/2020-21.

%%%%%%%%%%%%%%%%%%%%%%%%%%%%%%%%%%%%%%%%%%%%%%%%%%
\section*{Data Availability}

The simulated data underlying this work will be shared upon reasonable request to the corresponding author. 



%\nocite{*}
\bibliographystyle{mnras}
\bibliography{references}

%%%%%%%%%%%%%%%%% APPENDICES %%%%%%%%%%%%%%%%%%%%%

%\appendix

%%%%%%%%%%%%%%%%%%%%%%%%%%%%%%%%%%%%%%%%%%%%%%%%%%

% Don't change these lines
\bsp	% typesetting comment
\label{lastpage}
\end{document}